\documentclass[11pt]{article}
\usepackage[english]{babel}
\usepackage{cite}
\usepackage{amssymb,amsmath,mathrsfs}

\newtheorem{remark}{Remark}
\newtheorem{definition}{Definition}
\newtheorem{theorem}{Theorem}

\newcommand{\dequm}{\mathscr{D}}
\newcommand{\qum}{\mathscr{Q}}

\newcommand{\kerne}{\mathrm{Ker}\hspace{0.3mm}}
\newcommand{\ran}{\mathrm{Ran}\hspace{0.3mm}}

\newcommand{\hame}{\nu_G}
\newcommand{\esup}{\mbox{$\hame$-ess sup}}
\newcommand{\fphi}{\check{\varphi}}
\newcommand{\spb}{\mathsf{L\hspace{0.2mm}P}}
\newcommand{\spbn}{\mathsf{L\hspace{0.2mm}P\hspace{-0.5mm}}_n}
\newcommand{\norinf}{\|_\infty}
\newcommand{\nordu}{\|_2}

\newcommand{\Hstar}{\mathrm{H}^\ast\hspace{-0.5mm}}
\newcommand{\cinvo}{\hspace{0.2mm}\mathsf{I}\hspace{0.3mm}}
\newcommand{\sinvo}{\hspace{0.2mm}\mathsf{J}\hspace{0.3mm}}

\newcommand{\starp}{\hspace{-0.3mm}\circledast\hspace{-0.3mm}}
\newcommand{\stp}{\hspace{-0.3mm}\star\hspace{-0.3mm}}

\newcommand{\wfsn}{\mathsf{L\hspace{0.2mm}W}_n}
\newcommand{\pwfsn}{\mathsf{W}_n}
\newcommand{\pnwfsn}{\breve{\mathsf{W}}_n}

\newcommand{\copd}{\mathsf{P}}
\newcommand{\cpdn}{\mathsf{P\hspace{-0.5mm}}_n}
\newcommand{\cpdon}{\breve{\mathsf{P}\hspace{-0.5mm}}_n}
\newcommand{\lispn}{\mathsf{LQ}_n}
\newcommand{\qpdn}{\mathsf{Q}_n}
\newcommand{\qpdon}{\breve{\mathsf{Q}}_n}

\newcommand{\qpt}{\hspace{0.2mm}\mathcal{Q}\hspace{0.3mm}}
\newcommand{\obse}{\mathcal{A}}
\newcommand{\qua}{\hspace{0.2mm}\mathcal{Q}\hspace{0.2mm}}

\newcommand{\ccc}{\mathbb{C}}
\newcommand{\de}{\mathrm{d}}
\newcommand{\eee}{\mathrm{e}}
\newcommand{\dqdpn}{\de^n\hspace{-0.4mm}q\hspace{0.6mm}\de^n\hspace{-0.2mm}p}
\newcommand{\dqdpp}{\de^n\hspace{-0.4mm}q^\prime\hspace{0.2mm}\de^n\hspace{-0.2mm}p^\prime}
\newcommand{\dez}{\de z}
\newcommand{\dezp}{\de z^\prime}
\newcommand{\intrn}{\int_{\mathbb{R}^n}}
\newcommand{\dxn}{\;\mathrm{d}^n x\hspace{0.4mm}}
\newcommand{\facn}{{\frac{1}{(2\pi)^n}}}

\newcommand{\phasp}{\mathbb{R}^n\hspace{-0.4mm}\times\mathbb{R}^n}
\newcommand{\lrrn}{\mathrm{L}^2(\phasp)}
\newcommand{\lrrnco}{\mathrm{L}^2(\phasp,(2\pi)^{-n}\dqdpn\hspace{0.3mm};\mathbb{C})}
\newcommand{\rea}{(\phasp ; \mathbb{R})}
\newcommand{\dug}{\widehat{G}}
\newcommand{\cm}{\mathrm{CM\hspace{0.3mm}}}
\newcommand{\prm}{\mathrm{PM\hspace{0.3mm}}}
\newcommand{\pd}{\chi}
\newcommand{\lug}{\mathrm{L}^{1\hspace{-0.3mm}}(G)}
\newcommand{\ldrn}{\mathrm{L}^{2\hspace{-0.3mm}}(\mathbb{R}^n)}
\newcommand{\lurrn}{\mathrm{L}^{1\hspace{-0.3mm}}(\phasp)}
\newcommand{\tlurrn}{\mathcal{F}\hspace{0.3mm}\mathrm{L}^1\hspace{-0.3mm}(\phasp)}
\newcommand{\linfg}{\mathrm{L}^{\hspace{-0.4mm}\infty\hspace{-0.3mm}}(G)}
\newcommand{\linfrrn}{\mathrm{L}^{\hspace{-0.4mm}\infty\hspace{-0.3mm}}(\phasp)}

\newcommand{\invo}{^{\ast\hspace{-0.4mm}}}
\newcommand{\ccom}{\mathsf{C}_{\mathsf{c}}}
\newcommand{\gj}{g_j^{-1}}
\newcommand{\gk}{g_k^{\phantom{1}}}

\newcommand{\tr}{\mathrm{tr}}

\newcommand{\cast}{\mathrm{C}^\ast}

\newcommand{\tmu}{\widetilde{\mu}}

\newcommand{\zj}{z_j^{\phantom{\ast}}}
\newcommand{\zk}{z_k^{\phantom{\ast}}}

\newcommand{\coj}{\overline{c_j}}
\newcommand{\cok}{c_k}

\newcommand{\tvarrho}{\widetilde{\varrho}}

\newcommand{\clat}{\pd_t}
\newcommand{\clas}{\pd_s}
\newcommand{\clats}{\pd_{t+s}}
\newcommand{\claz}{\pd_0}

\newcommand{\csg}{\mathfrak{C}}
\newcommand{\hcsg}{\hat{\mathfrak{C}}}

\newcommand{\convo}{\hspace{-0.3mm}\circledcirc\hspace{-0.3mm}}

\newcommand{\qp}{{(q,p)}}

\newcommand{\disp}{\exp\!\left(\ima(p\cdot\hq-q\cdot\hp)\right)}
\newcommand{\hq}{{\hat{q}}}
\newcommand{\hp}{{\hat{p}}}
\newcommand{\tq}{\tilde{q}}
\newcommand{\tp}{\tilde{p}}

\newcommand{\fs}{\hat{\mathcal{F}}_{\hspace{-0.6mm}\mbox{\rm \tiny sp}}^{\phantom{x}}}
\newcommand{\fsy}{\hat{\mathcal{F}}_{\hspace{-0.6mm}\mbox{\rm \tiny sp}}}

\newcommand{\intrrn}{\int_{\phasp}\hspace{-1mm}}

\newcommand{\hs}{\mathcal{B}_2(\mathcal{H})}
\newcommand{\hsr}{\mathcal{B}_2(\ldrn)}
\newcommand{\ddu}{d_U^{\phantom{1}}}
\newcommand{\ee}{\mathrm{e}}

\newcommand{\tre}{\hspace{0.3mm}}
\newcommand{\quattro}{\hspace{0.4mm}}
\newcommand{\cinque}{\hspace{0.5mm}}
\newcommand{\sei}{\hspace{0.6mm}}
\newcommand{\sette}{\hspace{0.7mm}}
\newcommand{\otto}{\hspace{0.8mm}}
\newcommand{\nove}{\hspace{0.9mm}}
\newcommand{\dieci}{\hspace{1mm}}
\newcommand{\mtre}{\hspace{-0.3mm}}

\newcommand{\mcinque}{\hspace{-0.5mm}}
\newcommand{\msei}{\hspace{-0.6mm}}

\newcommand{\errep}{\mathbb{R}^{\mbox{\tiny $+$}}}

\newcommand{\opa}{\hat{A}}
\newcommand{\opb}{\hat{B}}

\newcommand{\cz}{\mathsf{C}_0}

\newcommand{\hh}{\mathcal{H}}

\newcommand{\hrho}{\hat{\rho}}

\newcommand{\mut}{\mu_t}

\newcommand{\muno}{\mu_1}
\newcommand{\mdue}{\mu_2}

\newcommand{\ima}{\mathrm{i}}

\newcommand{\two}{\mathcal{T}}

\newcommand{\trcr}{\mathcal{B}_1(\ldrn)}
\newcommand{\trcrp}{\mathcal{B}_1(\ldrn)^{\hspace{-0.3mm}\mbox{\tiny $+$}}}

\newcommand{\intG}{\int_G}

\newcommand{\urep}{{U\hspace{-0.5mm}\vee\hspace{-0.5mm} U}}

\newcommand{\unimu}{{\mut[\urep]}}

\newcommand{\emme}{\mathsf{m}\hspace{0.3mm}}
\newcommand{\modu}{\Delta_G}
\newcommand{\ldg}{\mathrm{L}^2(G)}

\begin{document}

\title{Playing with functions of positive type, classical and quantum}

\author{Paolo Aniello$^{1,2}$   \vspace{2mm}
\\
\small \it   $^1$Dipartimento di Fisica, Universit\`a di Napoli ``Federico II'',  \\
\small \it   Complesso Universitario di Monte S.\ Angelo, via Cintia, 80126 Napoli, Italy \vspace{1mm}
\\
\small \it   $^2$Istituto Nazionale di Fisica Nucleare, Sezione di Napoli,  \\
\small \it   Complesso Universitario di Monte S.\ Angelo, via Cintia, 80126 Napoli, Italy}

\date{}

\maketitle

\begin{abstract}
\noindent  A function of positive type can be defined as a
positive functional on a convolution algebra of a locally
compact group. In the case where the group is abelian,
by Bochner's theorem a function of positive type is, up
to normalization, the Fourier transform of a probability
measure. Therefore, considering the group of translations
on phase space, a suitably normalized phase-space function of positive type
can be regarded as a realization of a classical state. Thus,
it may be called a \emph{function of classical positive type}.
Replacing the ordinary convolution on phase space with the \emph{twisted}
convolution, one obtains a noncommutative algebra of functions
whose positive functionals we may call \emph{functions of quantum
positive type}. In fact, by a quantum version of Bochner's theorem,
a \emph{continuous} function of quantum positive type is, up
to normalization, the (symplectic) Fourier transform of
a Wigner quasi-probability distribution; hence, it can be
regarded as a phase-space realization of a quantum state.
Playing with functions of positive type --- classical and quantum ---
one is led in a natural way to consider a class of semigroups of operators, the
\emph{classical-quantum semigroups}. The physical meaning
of these mathematical objects is unveiled via quantization,
so obtaining a class of quantum dynamical semigroups
that, borrowing terminology from quantum information science, may be
called \emph{classical-noise semigroups}.
\end{abstract}


\section{Introduction}
\label{introduction}

As is well known, theoretical physics and mathematics are inextricably
intertwined sciences. A striking example of this fruitful relationship
is the profound role that the theory of groups and of group representations
plays in quantum mechanics and quantum field theory, and, conversely, the
impulse that the former subject has received due to its interaction
with the latter~\cite{Weyl,Wigner-book,Mackey1,Mackey2,Raja,Raja-bis,FollandQFT}.
In this context, a central position --- historically and conceptually ---
is certainly held by the phase-space approach to quantum theory, a topic that
has developed into a vast and interesting research field
thanks to the pioneering work of outstanding scientists like Weyl, Wigner,
Groenewold and Moyal~\cite{Weyl,Wigner,Groenewold,Moyal}, at an early stage,
and to the efforts of several other researchers up to this day;
see, e.g.,~\cite{Hillery,FollandHAPS,Ali,Schleich,Dodonov,Aniello-coher,AnielloFT,AnielloSP},
and references therein.

The present paper is aimed at providing a simple example of how
a mathematical notion arising in the realm of (abstract) \emph{harmonic analysis}
--- essentially, the theory of topological groups, of their representations
and of suitable spaces of functions on such groups ---
assumes a precise meaning when considered from the point of view of
classical statistical mechanics or of phase-space quantum mechanics,
and may give rise to new insights and applications.

A fundamental concept in harmonic analysis is that of \emph{function of positive type}
on a locally compact group~\cite{Godement,Folland-AA}. In the case where the group is abelian,
by a classical theorem of Bochner~\cite{Rudin,Klenke} such a function can be regarded as
the Fourier transform of a probability measure --- a so-called
\emph{characteristic function} --- up to normalization.
In classical statistical mechanics, probability measures play the role
of physical states. Thus, it turns out that one can realize such a state,
alternatively, by a function of positive type on phase space.
The latter realization is often more convenient to deal with, since it relies on
ordinary $\ccc$-valued functions, rather than on quite
abstract objects like measures.

In the standard formulation of quantum mechanics, the physical states
are realized as density operators, namely, as normalized, positive trace class
operators. In this language, there is no direct link with the notion
of function of positive type. However, from an algebraic point of view,
states are positive (linear) functionals on the $\ast$-algebra of observables
(the $\cast$-algebra algebra of bounded operators)~\cite{Emch,Strocchi}.
This is analogous to the classical case, where the complex (Radon) measures~\cite{FollandRA}
on phase space provide a realization of the bounded functionals
on the commutative algebra of continuous complex functions
vanishing at infinity. In this setting, the states are the normalized positive functionals
--- namely, the probability measures --- while
the observables are the real functions (the selfadjoint part of the $\ast$-algebra of observables).
Alternatively, as already mentioned, one can
realize the physical states as normalized functions of positive type on phase space,
that can be regarded as functionals on the Banach space of integrable functions,
which is an algebra with respect to convolution.

The analogy between the classical and the quantum case becomes much deeper, however,
by considering the phase-space formulation of quantum mechanics. In this approach,
states are realized as quasi-probability distributions (the Wigner functions)
and the (symplectic) Fourier transform of these distributions ---
the \emph{quantum characteristic functions} --- can also be regarded
as \emph{suitably} normalized positive functionals on a certain noncommutative
$\ast$-algebra of square integrable functions.
We call (all) the positive functionals on this algebra \emph{functions of quantum positive type}.
The relation between the intrinsic characterization of functions of quantum positive type
and the property of being a quantum characteristic function is
a quantum counterpart of Bochner's theorem~\cite{Kastler,LMS1,LMS2}
(also see~\cite{Narcowich1,Narcowich2}).

Beside these remarkable analogies between the classical and the quantum case,
there are a few noteworthy differences. Whereas in the classical setting
functions of positive type are automatically continuous, in the quantum setting
continuity is precisely the condition that allows one to single out, up to normalization,
the quantum characteristic functions among all functions of
quantum positive type. Moreover, in the classical case functions of
positive type are normalized in the sense of functionals. In the
quantum case, a different kind of normalization criterion
(for the continuous functions of quantum positive type) has to be applied
for selecting the physical states and, for these states, normalization in the
sense of functionals has to be interpreted as the square root of \emph{purity}~\cite{Lupo}.
The two normalization criteria coincide precisely for \emph{pure} states.

The framework of functions of positive type can also be regarded as
a common arena where classical and quantum states can be put all together
in the same game. Indeed, probability measures on phase space form a semigroup,
with respect to convolution, and, accordingly,
functions of positive type form a semigroup with respect to point-wise multiplication.
In this context, natural objects are a convolution semigroup
of probability measures and, associated with this, a multiplication
semigroup of (normalized) positive definite functions. On the
other hand, one can show that the point-wise product of a
function of positive type by a continuous function of \emph{quantum}
positive type is again a function of the \emph{latter} type.
These ingredients allow one to obtain a semigroup of operators ---
a so-called \emph{classical-quantum semigroup}~\cite{AnielloCQS} --- acting
in the Banach space generated by the continuous functions of
quantum positive type, out of a multiplication
semigroup of positive definite functions. At this point,
it is natural to wonder whether such a semigroup of
operators, which arises when playing with functions of positive type,
classical and quantum, has any physical meaning.

By quantizing a classical-quantum semigroup, one obtains a
new semigroup of operators acting in the Banach space of
trace class operators. This semigroup of operators turns out
to be a \emph{quantum dynamical semigroup}, namely, it
describes the evolution of an open quantum system~\cite{Holevo,Breuer}.
The quantized counterpart of a classical-quantum semigroup
belongs, in particular, to the class of
\emph{twirling semigroups}~\cite{Aniello1,Aniello2,Aniello3}, and has a
precise role in the context of quantum information science.
Because of this role, we will call it a \emph{classical-noise semigroup}.

The paper is organized as follows. In sect.~\ref{posdef}, we introduce
the notions of function of classical (subsect.~\ref{ccase})
and quantum (subsect.~\ref{qcase}) positive type,
and we discuss their physical meaning. Next, in sect.~\ref{semigroupops},
we show that associated with every multiplication semigroup of
functions of positive type is associated a semigroup of operators, i.e.,
a classical-quantum semigroup. Via quantization,
one finds out that a classical-quantum semigroup is mapped into
a quantum dynamical semigroup; see sect.~\ref{quan-dequa}. Finally,
in sect.~\ref{conclus}, a few conclusions are drawn.


\section{Functions of positive type: classical and quantum}
\label{posdef}

In this section, we will establish some basic facts concerning
functions of positive type on phase space and their quantum counterpart,
and discuss the relevance of these functions in classical (statistical)
mechanics and in quantum mechanics. Usually, we will feel
free to mention and use well known results from functional analysis and abstract
harmonic analysis that can be found in standard references; see,
e.g.,~\cite{Godement,Folland-AA,Rudin} for functions of positive type,
~\cite{Folland-AA,Rudin,FollandRA} for Fourier analysis on abelian groups
and complex measures, \cite{Strocchi,Murphy} for $\ast$-algebras
and~\cite{FollandHAPS,Ali,AnielloFT,AnielloSP} for Wigner distributions,
the associated quantum characteristic functions and the twisted convolution.
Explicit reference to the literature will be made for nonstandard facts only.

\subsection{The classical case}
\label{ccase}

Given a locally compact group $G$ (precisely, we will assume that it is a locally compact,
second countable, Hausdorff topological group), let $\lug$ be the Banach space
of $\ccc$-valued functions on $G$, integrable w.r.t.\ the left Haar measure $\hame$.
As is well known, $\lug$ --- endowed with the convolution product $(\cdot)\convo(\cdot)$,
\begin{equation}
(\varphi_1\convo\varphi_2)(g):= \intG \varphi_1(h)\tre\varphi_2(h^{-1}g) \; \de\hame(h),
\end{equation}
and with the involution $\cinvo\colon\varphi\mapsto\varphi\invo$,
\begin{equation}
\varphi\invo(g):=\modu(g^{-1})\cinque \overline{\varphi(g^{-1})},
\end{equation}
with $\modu$ denoting the modular function on $G$ ---
becomes a Banach $\ast$-algebra $\big(\lug, \convo, \cinvo\big)$.

\begin{definition} \label{defclpoty}
A positive, bounded linear functional on the Banach $\ast$-algebra $\big(\lug, \convo, \cinvo\big)$,
realized as a function in the Banach space of $\hame$-essentially bounded functions $\linfg$,
is called a \emph{function of positive type} on $G$. Namely, a function
$\pd$ on $G$ is said to be of positive type if it belongs to $\linfg$ and
\begin{equation} \label{clapo}
\intG \pd (g) \cinque (\varphi\invo \convo \varphi) (g) \; \de\hame(g)\ge 0,
\end{equation}
for all $\varphi \in \lug$.
\end{definition}

\begin{remark}
{\rm
By our previous assumptions, $G$ is $\sigma$-compact and $\hame$ is $\sigma$-finite,
so the Banach space $\linfg$ of essentially bounded functions on $G$, defined
in the usual way~\cite{FollandRA}, can be identified with the dual space of $\lug$,
and certain technicalities regarding the (re-)definition of $\linfg$
--- see, e.g., sect.~{2.3} of~\cite{Folland-AA} --- are not necessary in this case.
}
\end{remark}

A function of positive type $\pd\in\linfg$ agrees $\hame$-almost everywhere with a (bounded)
continuous function\footnote{As in Definition~\ref{defclpoty}, with a standard abuse, here and in the following
we do not make a distinction between an element of $\linfrrn$, which is an equivalence class
of functions, and a representative in this class.} and
\begin{equation} \label{nori}
\|\pd\norinf := \esup_{g\in G} \sei |\pd(g)|=\pd(e),
\end{equation}
with $e$ denoting the identity in $G$ and $\pd(e)$ the value at $e$ of the `continuous version' of $\pd$.

For a bounded continuous function $\pd\colon G\rightarrow\ccc$ the following facts are equivalent:
\begin{description}

\item[P1)] $\pd$ is of positive type;

\item[P2)] $\pd$ satisfies condition~{(\ref{clapo})}, for all $\varphi \in \ccom(G)$
(the linear space of continuous $\ccc$-valued functions on $G$, with compact support);

\item[P3)] $\pd$ satisfies the condition
\begin{equation} \label{clapo-bis}
\intG \intG \pd (g^{-1}h) \cinque \overline{\varphi(g)}\quattro \varphi(h) \; \de\hame(g)\quattro\de\hame(h)\ge 0,
\end{equation}
for all $\varphi \in \ccom(G)$;

\item[P4)] $\pd$ is a \emph{positive definite function}, i.e.,
\begin{equation} \label{pdfncts}
\sum_{j,k}\pd(\gj\gk)\otto \coj\sei\cok \ge 0,
\end{equation}
for every finite set $\{g_1,\ldots,g_m\}\subset G$ and arbitrary complex numbers $c_1,\ldots,c_m$.

\end{description}

Note that the equivalence between the second and the third point above
is obtained by a simple change of variables in the integrals, and
relation~{(\ref{pdfncts})} can be regarded as a discretized version of~{(\ref{clapo-bis})}.

Let us now focus on the case where $G$ is abelian (in particular, a vector group).
We will denote by $\dug$ the Pontryagin dual of $G$ and by $\cm(\dug)$ the Banach space
of complex Radon measures on $\dug$.
In this case, according to Bochner's theorem, we can add a further item
to the previous list of equivalent properties:
\begin{description}

\item[P5)] $\pd$ is the Fourier (i.e., Fourier-Stieltjes) transform of a positive measure $\mu\in\cm(\dug)$.

\end{description}

In the case where the positive measure $\mu\in\cm(\dug)$ is a probability measure --- $\mu(\dug)=1$ ---
this normalization condition translates into $\pd(0)=1$
(with $0$ denoting the identity in the abelian group $G$) and, taking into account
relation~{(\ref{nori})}, the latter condition is nothing but
the normalization of $\pd$ regarded as a functional on $\lug$.
In the context of probability theory,
$\pd$ is usually called the \emph{characteristic function}
associated with $\mu$~\cite{Klenke}.

If $G$ is the group of translations on the $(n+n)$-dimensional phase space
--- i.e., the vector group $\phasp$ --- we can
obviously identify $\dug$ with $G$ itself, and use the
\emph{symplectic} Fourier transform~\cite{FollandHAPS}, instead of the ordinary one.

The physical meaning and relevance of functions of positive type become
evident as soon as one considers that probability measures on phase space
play the role of physical \emph{states} in classical statistical mechanics;
see, e.g.,~\cite{Strocchi}.
Indeed, from the algebraic point of view a classical state is a normalized positive
functional on the $\cast$-algebra of (classical) observables. Taking into account
Gelfand theory of commutative $\cast$-algebras~\cite{Murphy},
according to which such an algebra is isomorphic to  an algebra of continuous functions
vanishing at infinity, a natural choice for the algebra of classical observables on phase space
is $\cz(\phasp)$, the Banach space of continuous $\ccc$-valued functions on $\phasp$ vanishing at infinity,
which is a $\cast$-algebra if endowed with the point-wise product (the \emph{true} observables forming
the selfadjoint part $\cz\rea$ of the whole algebra). The dual space of $\cz(\phasp)$ is
$\cm(\phasp)$, the space of complex Radon measures on $\phasp$, and the states
on the $\cast$-algebra $\cz(\phasp)$ are precisely the probability measures on
$\phasp$. Clearly, the expectation value of an observable $f\in\cz\rea$
in the state $\mu\in\cm(\phasp)$ is provided by the expression
\begin{equation} \label{clev}
\langle f\rangle_{\mu} =  \intrrn f(q,p)\, \de\mu(q,p) .
\end{equation}

Unfortunately, probability measures are in general rather abstract objects to
deal with, so that it is often useful to replace a classical state $\mu\in\cm(\phasp)$ with its
symplectic Fourier transform:
\begin{equation} \label{ftmes}
\tmu(q,p):=\intrrn \eee^{\ima (q\cdot p^\prime - p\cdot q^\prime)}\; \de\mu(q^\prime,p^\prime).
\end{equation}
Moreover, as is well known, the (symplectic) Fourier transform maps $\lurrn$,
continuously and injectively,
into a dense linear subspace $\tlurrn$ of $\cz(\phasp)$:
\begin{equation}
\varphi\mapsto \fphi, \ \ \
\fphi(q,p) := \intrrn \varphi(q^\prime, p^\prime)\otto \eee^{\ima (q^\prime\mcinque\cdot p - p^\prime\mcinque\cdot q)}\;
\dqdpp .
\end{equation}
Then, by Fubini's theorem, for every observable $f\in\tlurrn$
the expectation value $\langle f\rangle_{\mu}$ can be expressed in
the following form:
\begin{equation}
\langle f\rangle_{\mu} = \intrrn \varphi (q,p) \sei \pd (q,p)\; \dqdpn =: \langle \varphi,\pd\rangle \, ,
\ \ \ \mbox{with $\fphi= f$ and $\chi=\tmu$}.
\end{equation}
Here $\langle \varphi,\pd\rangle$ should be regarded as the pairing
between $\varphi\in\lurrn$ and $\pd\in\linfrrn$ (a function of positive type).

\begin{remark}
{\rm
The condition that $f\in\tlurrn$ be an observable --- $f=\overline{f}$ ---
corresponds to the condition that $\varphi$ be \emph{selfadjoint}: $\varphi=\cinvo\tre\varphi$;
namely, that $\varphi(q,p)=\overline{\varphi(-q,-p)}$.
}
\end{remark}

\begin{remark}
{\rm
The pairing between $\cz(\phasp)$ and $\cm(\phasp)$ can be completely restored
considering that the $\cast$-algebra $\cz(\phasp)$ is isomorphic, via the Fourier transform,
to the group $\cast$-algebra $\cast (\phasp)$ (the universal $\cast$-completion of $\lurrn$), and the
dual of the Banach space $\cast (\phasp)$ can be identified with
the subspace $\spbn\equiv\spb(\phasp)$ of $\linfrrn$ consisting of all complex linear
superpositions of continuous functions of positive type on $\phasp$; see~\cite{Eymard}.
}
\end{remark}

We can summarize the picture outlined above as follows:
\begin{itemize}

\item The notion of function of positive type is group-theoretical and
can be defined for any locally compact group, giving rise to various
equivalent characterizations. In the case of an abelian group, a function of
positive type can be characterized, up to normalization, as the Fourier transform
of a probability measure.

\item In the standard approach to classical statistical mechanics,
the physical states are realized as probability measures on phase space,
while the observables form the selfadjoint part of the $\cast$-algebra $\cz(\phasp)$.

\item Alternatively, one can undertake a \emph{characteristic function approach}
by replacing probability measures with the corresponding normalized functions
of positive type (characteristic functions). In this approach, the standard
observables are (densely) replaced with the functions belonging to the
selfadjoint part of the Banach $\ast$-algebra $\lurrn$.

\item Whereas the dual space of $\lurrn$ is $\linfrrn$, the \emph{positive} functionals
on $\lurrn$ are precisely the (continuous) functions of positive type on phase space.
Therefore, in the characteristic function approach $\lurrn$ may be thought of as the
algebra of observables \emph{tout court} (instead of the group $\cast$-algebra $\cast (\phasp)$).

\end{itemize}

Continuous functions of positive type form a convex cone $\cpdn\equiv\copd(\phasp)$ in
$\linfrrn$, and characteristic functions --- i.e., normalized continuous functions of positive type ---
form a convex subset $\cpdon$ of $\cpdn$. Since, as it will be shown in subsect.~\ref{qcase}, there is
a quantum analogue of functions of positive type on $\phasp$, the elements of $\cpdn$  will be sometimes referred to as
\emph{functions of classical positive type}.

\subsection{The quantum case}
\label{qcase}

In the standard formulation of quantum mechanics \emph{normal states}~\cite{Emch}
are realized as \emph{density operators}, namely, as
normalized, positive trace class operators in a certain Hilbert space.
Within this formalism, there is no \emph{direct} connection between a physical state
and anything analogous to the notion of function of positive type.
On the other hand, a profound link of this kind does emerge
if one adopts a phase-space approach to quantum mechanics.
Whereas various phase-space formalisms have been proposed in the literature,
see e.g.~\cite{Husimi,Kano,Sudarshan,Mehta,Glauber1,Glauber2,Glauber3,Mancini,AnielloFT},
the prototype is certainly the remarkable approach developed by
Weyl, Wigner, Groenewold and Moyal~\cite{Weyl,Wigner,Groenewold,Moyal}
--- which includes a suitable quantization-dequantization scheme
and the notion of star product of functions --- approach
that may be called the \emph{WWGM formulation} of quantum mechanics.

In the WWGM formulation, a \emph{pure} state
$\hrho_\psi=|\psi\rangle \langle\psi |$, $\psi\in\ldrn$ ($\|\psi\|=1$),
is replaced with a function $\varrho_\psi\colon\phasp\rightarrow\ccc$
defined according to Wigner's classical prescription, i.e.,
\begin{equation}
\varrho_\psi (q,p) := \facn\intrn \eee^{-\ima p \cdot x}\,
\psi\!\left(q-\frac{x}{2}\right)^*
\psi\!\left(q+\frac{x}{2}\right)\dxn ,
\end{equation}
where we have set $\hbar=1$.
This definition, by means of the spectral decomposition of a positive trace class
operator, extends in a natural way to any density operator in $\ldrn$ and then,
by taking (finite) linear superpositions, to every trace class operator.
The phase-space functions associated with (normal) states are
usually called \emph{Wigner functions}.

By this construction, one obtains a separable complex Banach
space of functions that will be denoted by $\wfsn$
(the image via the Wigner map, see sect.~\ref{quan-dequa} {\it infra},
of the Banach space of trace class operators in $\ldrn$).
This linear space contains a convex cone $\pwfsn$, formed by those functions that are associated with
\emph{positive} trace class operators in $\lrrn$, and $\pwfsn$ contains the convex set
$\pnwfsn$ formed by the Wigner functions. Within $\pwfsn$, the Wigner functions are precisely
those functions satisfying the normalization condition
\begin{equation} \label{norcon}
\lim_{r\rightarrow +\infty} \int_{|q|^2+|p|^2<r} \mcinque  \varrho(q,p) \; \dqdpn = \tr(\hrho) =1,
\end{equation}
see~\cite{Daubechies-bis}, where $\varrho\in\pnwfsn$ is the
function canonically associated with a certain state $\hrho$.

A Wigner function $\varrho$ is real but, in general,
it is not a genuine probability distribution, since it may assume negative values;
moreover, it is square integrable but, in general, \emph{not} integrable~\cite{Daubechies-bis},
fact that is taken into account in the lhs of~{(\ref{norcon})}.
Nevertheless, one can express the expectation value of an observable $\hat{A}$
in the state $\hrho$ --- $\langle \hat{A}\rangle_{\hrho} =  \tr (\hat{A}\,\hrho\tre)$) ---
as an integral on phase space,
\begin{equation} \label{expeval}
\langle \hat{A}\rangle_{\hrho} = \intrrn \msei \alpha(q,p)\, \varrho(q,p) \;
\dqdpn ,
\end{equation}
where $\alpha$ is a real function associated with the operator $\hat{A}$.
Formula~{(\ref{norcon})} and the normalization condition~{(\ref{expeval})}
explain why the Wigner function $\varrho$
is often called a \emph{quasi-probability distribution}.

\begin{remark}
{\rm
Actually, not all the observables can be realized as ordinary functions in the WWGM formulation.
In general, they form a suitable class of distributions (generalized functions), see~\cite{Daubechies}.
This fact, however, will not play any role in the following.
}
\end{remark}

As in the classical setting, also in the quantum case one can replace
a Wigner quasi-probability distribution with its symplectic Fourier
transform. Precisely, taking into account the fact that a Wigner function
is square integrable (but, in general, not integrable), one should use
the symplectic Fourier-Plancherel operator $\fs$, which is the
selfadjoint unitary operator in $\lrrn$ determined by
\begin{equation} \label{detsfp}
\big(\fs f\big)(q,p)=\frac{1}{(2\pi)^n}\intrrn f(q^\prime,p^\prime)\,
\ee^{\ima (q\cdot p^\prime - p\cdot q^\prime)}\; \dqdpp ,
\end{equation}
for all $f\in\lurrn\cap\lrrn$.
Then, the space $\wfsn$ is mapped onto a dense subspace $\lispn$
of $\lrrn$,
\begin{equation}
\lispn:= \fs \sei \wfsn ,
\end{equation}
and the convex cone $\pwfsn\subset\wfsn$ is mapped onto
a convex cone $\qpdn\subset\lispn$. Each function in
$\lispn$ agrees almost everywhere (w.r.t.\ Lebesgue measure)
with a continuous function. Actually, it will often be convenient, in the following,
to regard $\lispn$ as a linear space of continuous functions.

By analogy with the classical case, we may call a function $\tvarrho$ in
$\qpdn$, defined by
\begin{equation} \label{wigetcha}
\tvarrho := (2\pi)^n \cinque \fs\sei\varrho , \ \ \ \varrho\in\pnwfsn,
\end{equation}
the \emph{quantum characteristic function} associated with the
quasi-probability distribution $\varrho$. The factor $(2\pi)^n$ appearing in~{(\ref{wigetcha})}
is chosen in such a way that the quantum characteristic functions, similarly to the classical case,
are those functions in $\qpdn$ satisfying the normalization condition
\begin{equation} \label{qposba}
\tvarrho(0)=1 ,
\end{equation}
with $0\equiv(0,0)$ denoting the origin in $\phasp$.
Thus, the quantum characteristic functions form a convex subset
$\qpdon=(2\pi)^n \cinque \fs\sei\pnwfsn$ of $\lispn$.

A natural task in the WWGM approach is to provide a \emph{fully} self-consistent formulation
of quantum theory, task that obviously includes the problem of characterizing \emph{intrinsically}
the convex set of physical states $\pnwfsn$ or, equivalently, the
convex set $\qpdon$ of quantum characteristic functions. The analysis of this problem
leads in a natural way to the notion of \emph{function of quantum positive type}.

As in the classical setting, we first consider a $\ast$-algebra of functions,
and then define the functions of positive type as suitable functionals
on this algebra. To this aim, we recall that the Hilbert space $\lrrn$
becomes a $\ast$-algebra --- precisely, a $\Hstar$-algebra~\cite{AnielloSP,Rickart} ---
once endowed with the \emph{twisted convolution} $(\cdot)\starp(\cdot)$,
\begin{equation} \label{deftc}
\big(\obse_1\starp \obse_2\big)(q,p)
=  \frac{1}{(2\pi)^n}\intrrn \obse_1(q^\prime,p^\prime) \,
\obse_2(q-q^\prime,
p-p^\prime)\otto\eee^{\frac{\ima}{2}(q\cdot p^\prime-p\cdot q^\prime)}
\, \dqdpp,
\end{equation}
$\obse_1, \obse_2\in\lrrn$, and with the involution $\sinvo\colon \obse\mapsto \obse\invo$,
\begin{equation} \label{forinvo}
\obse\invo(q,p) := \overline{\obse(-q,-p)},\ \ \ \obse\in\lrrn  .
\end{equation}
Note that involution $\sinvo$ is formally identical to that arising in the classical case,
but it acts in a different space. An element $\obse$ of $\lrrn$ is said to be
\emph{selfadjoint} if $\obse=\sinvo\tre\obse$.
We will implicitly use the fact that the twisted convolution
of a pair of functions in $\ccom(\phasp)$ belongs to this space too;
hence, $\ccom(\phasp)$ is a $\ast$-subalgebra of $\lrrn$.
One can show, moreover, that $\lrrn\starp\lrrn=\lispn$ and $\sinvo\nove\lispn=\lispn$.
Thus, $\lispn$ is another $\ast$-subalgebra of $\lrrn$.
Incidentally, we mention that the twisted convolution allows one to describe the norm with respect to which
$\lispn$ becomes a Banach space isomorphic to $\wfsn$.
However, since this will not be relevant for our purposes, we will
not insist on this point. It is instead worth anticipating that the twisted convolution can be regarded
as an expression of the product of operators in terms of phase-space functions, see sect.~\ref{quan-dequa},
and with reference to this fact it is often called a \emph{star product}~\cite{AnielloFT,AnielloSP}.

Clearly, the dual of the Hilbert space $\lrrn$ can be identified with
$\lrrn$ itself; hence, the functions of quantum positive type should be square integrable.

\begin{definition}
A positive, bounded linear functional on the $\Hstar$-algebra $\big(\lrrn, \starp, \sinvo\big)$, implemented by
a square integrable function on $\phasp$, is called a \emph{function of quantum positive type}.
Otherwise stated, we say that a function $\qpt\colon\phasp\rightarrow\ccc$ is of
quantum positive type if it belongs to $\lrrn$ and
\begin{equation} \label{quapo}
\intrrn \mcinque\qpt(q,p)\sette (\obse\invo\starp\obse)(q,p)  \; \dqdpn \ge 0,
\end{equation}
for all $\obse\in\lrrn$.
\end{definition}

\begin{remark}
{\rm
Condition~{(\ref{quapo})} is equivalent to the following:
\begin{equation} \label{quapoalt}
\intrrn \mcinque\qpt(q,p)\sette \overline{(\obse\invo\starp\obse)(q,p)}  \; \dqdpn \ge 0,
\ \ \ \forall\quattro\obse\in\lrrn .
\end{equation}
This fact can be easily deduced from the relation
$\obse\invo\starp\obse= \overline{(\overline{\obse}\tre\starp(\overline{\obse})\invo)}$.
}
\end{remark}

This definition introduces a notion of function of positive type, but at this stage,
it is not evident why one should regard such an object as `quantum'. To clarify this point,
we will argue that, as in the classical case, a function of quantum positive type
can be characterized by means of various equivalent properties. One of these has a clear physical meaning.

It will be convenient, in the rest of this section, to adopt a shorthand notation
for points on phase space and set: $z\equiv (q,p)\in\phasp$,
$\dez\equiv \dqdpn$. The standard symplectic form will be denoted by $\omega$, i.e.,
\begin{equation}
\omega(z\tre ,\mtre z^\prime):= q\cdot p^\prime-p\cdot q^\prime .
\end{equation}

Using results from~\cite{Kastler,LMS1,LMS2,Narcowich2}, one can verify
the following facts. If a \emph{continuous} function $\qpt$ is of quantum positive type,
then it is bounded and
\begin{equation} \label{norqpt}
\|\qpt\norinf = \qpt(0) .
\end{equation}
Note that this relation is analogous to~{(\ref{nori})}. Moreover,
for a \emph{continuous} function $\qpt\colon\phasp\rightarrow\ccc$ the following properties
are equivalent:
\begin{description}

\item[Q1)] $\qpt$ is of quantum positive type;

\item[Q2)] $\qpt$ satisfies condition~{(\ref{quapo})}, for all $\obse \in \ccom(\phasp)$;

\item[Q3)] $\qpt$ satisfies the condition
\begin{equation} \label{quapo-bis}
\intrrn \intrrn \mcinque\qpt (z-z^\prime)\sei \overline{\obse(z^\prime)}\sei \obse(z)\otto
\eee^{\ima\tre \omega(z^\prime\mcinque ,\tre z)/2} \; \dez \sei \dezp \ge 0,
\end{equation}
for all $\obse \in \ccom(\phasp)$;

\item[Q4)] $\qpt$ is a \emph{quantum positive definite function}, i.e.,
\begin{equation} \label{qpdfncts}
\sum_{j,k}\qpt(\zk-\zj)\nove
\eee^{\ima\tre \omega(\zj\mtre,\tre\zk)/2}\; \coj\sei\cok \ge 0 ,
\end{equation}
for every finite set $\{z_1,\ldots,z_m\}\subset \phasp$ and arbitrary complex numbers $c_1,\ldots,c_m$;

\item[Q5)] $\qpt$ belongs to the convex cone $\qpdn\subset\lispn$, i.e.,
it is a multiple of the Fourier-Plancherel transform of a Wigner quasi-probability
distribution.

\end{description}

The last characterization of functions of quantum positive type in the above list
is clearly a quantum version of Bochner's theorem
(Kastler 1965~\cite{Kastler}, Loupias and Miracle-Sole 1966-1967~\cite{LMS1,LMS2}).
By this characterization, $\qpdn$ can be regarded as
the set of continuous functions of quantum positive type, and $\lispn$
as the complex vector space generated by finite linear superpositions of functions of this kind.
By relations~{(\ref{qposba})} and~{(\ref{norqpt})}, the convex set $\qpdon$
of quantum characteristic functions coincides with the set of continuous functions of
quantum positive type satisfying the normalization condition: $\|\qpt\norinf = \qpt(0)=1$.
For every $\qpt\in\qpdon$, the norm of $\qpt$ as a functional (namely, as an element of $\lrrn$)
verifies the relation
\begin{equation} \label{anticinor}
\|\qpt\nordu \le 1,
\end{equation}
for a suitable normalization of the Haar measure on $\phasp$ (i.e., a multiple of the Lebesgue measure);
see sect.~\ref{quan-dequa} {\it infra}.
This inequality is saturated if and only if $\qpt$ is a \emph{pure} state, fact that
will become clear in sect.~\ref{quan-dequa}, where it will be argued that the square
of $\|\qpt\nordu$ is nothing but the purity of the density operator associated with $\qpt$.

\begin{remark}
{\rm
Recall that, in the classical case, a phase-space function of positive type is
automatically continuous --- precisely, it agrees almost everywhere with a continuous function.
Thus, requiring continuity may be regarded as a way of removing the ambiguity
in the choice of an item among all functions implementing the same functional on $\lurrn$
(namely, among all functions giving rise to the same element of $\linfrrn$).
On the other hand, in the quantum case, continuity must be imposed in order
to single out, up to normalization, precisely the quantum characteristic functions.
}
\end{remark}

Comparing relations~{(\ref{clapo})}, {(\ref{clapo-bis})} and~{(\ref{pdfncts})}
with~{(\ref{quapo})}, {(\ref{quapo-bis})} and~{(\ref{qpdfncts})}, respectively, one
should note something more than a formal similarity. In the latter inequalities,
a major role is played by the function
\begin{equation}
(\phasp)\times(\phasp)\ni (z,z^\prime)\mapsto \eee^{\ima\tre \omega(z , \tre z^\prime)/2},
\end{equation}
which is a nontrivial multiplier for the group $\phasp$~\cite{Raja}, the multiplier
associated with the Weyl system and with the integrated form of canonical commutation relations~\cite{Aniello-WS},
whereas in inequalities~{(\ref{clapo})}, {(\ref{clapo-bis})} and~{(\ref{pdfncts})}
the trivial multiplier is implicitly involved. This is quite an impressive fact:
the symplectic structure that plays such a fundamental role in classical mechanics
emerges in a spontaneous way in the quantum setting, e.g., in condition~{(\ref{quapo-bis})}
satisfied by quantum characteristic functions.


\section{Playing with functions of positive type}
\label{semigroupops}

The \emph{convolution} $\muno\convo\mdue$ of a pair of positive measures $\muno,\mdue\in\cm(G)$
on a locally compact group $G$ ---
\begin{equation}
\intG \varphi(g) \; \de\muno\convo\mdue(g) := \intG \intG \varphi(gh) \; \de\muno(g)\tre\de\mdue(h), \ \ \
\varphi\in\ccom(G)
\end{equation}
--- is a positive measure in $\cm(G)$ too~\cite{Folland-AA}; in particular,
a probability measure if $\muno$ and $\mdue$ are both normalized. Endowed with
convolution the set $\prm(G)$ of probability measures on $G$ becomes a semigroup.
The identity of this semigroup is given by the
Dirac measure at the identity of $G$. If $G$ is abelian, to the convolution of probability measures corresponds,
via the Fourier(-Stieltjes) transform, the point-wise multiplication of characteristic functions~\cite{Rudin}.
By Bochner's theorem, it follows that the point-wise product $\pd_1\sei\pd_2$ of two continuous functions
of positive type $\pd_1$ and $\pd_1$ on $G$ is again a continuous function of positive type,
and clearly point-wise multiplication preserves normalization too.

Therefore --- considering now the case where $G=\phasp$ ---
endowed with the point-wise product of functions the set $\cpdon$
of normalized functions of classical positive type is a semigroup,
the identity being the function $\pd\equiv 1$. The Fourier transform
determines an algebraic isomorphisms between the semigroups $\prm(\phasp)$ and
$\cpdon$, which becomes a topological isomorphism if $\prm(\phasp)$ and
$\cpdon$ are endowed with the weak topology and the topology of
uniform convergence on compact sets, respectively (L\'evy's continuity theorem, see~\cite{Klenke}).
The latter topology turns out to coincide with the topology induced
on $\cpdon$ by the weak$^\ast$-topology of $\linfrrn$
(regarded as the dual of $\lurrn$), see~\cite{Raikov,Yoshizawa},
which is the natural one in this context.

What happens with the point-wise multiplication of a function of \emph{classical} positive type
by a continuous function of \emph{quantum} positive type?
\begin{theorem} \label{mainth}
The point-wise product $\pd\tre\qpt$ of a function $\pd\in\cpdn$ by a function $\qpt\in\qpdn$
belongs to $\qpdn$; in particular, it belongs to $\qpdon$ if $\pd$ and $\qpt$ are normalized.
\end{theorem}

\noindent {\it Proof:}  First observe that condition~{(\ref{pdfncts})}, for $G=\phasp$,
becomes
\begin{equation} \label{pdfncts-bis}
\sum_{j,k}\pd(\zk-\zj)\otto \coj\sei\cok \ge 0, \ \ \ z\equiv (q,p).
\end{equation}
By Schur's product theorem~\cite{Horn}, the Hadamard product
(i.e., the entrywise product) of two positive (semi-definite) matrices is positive too.
Applying this result to conditions~{(\ref{pdfncts-bis})} and~{(\ref{qpdfncts})}, that
for continuous functions on $\phasp$ are equivalent to the property of being of positive type
--- classical and quantum, respectively --- one readily proves the result.~$\square$

The previous result allows us to play with the point-wise
product of functions of positive type (classical or quantum).
Consider then a \emph{multiplication semigroup of functions of positive type}, i.e.,
a set $\{\clat\}_{t\in\errep}$ of normalized continuous functions of positive type
on $\phasp$ such that
\begin{equation}
\clat\otto\clas=\clats, \ t,s\ge 0, \ \ \
\claz \equiv 1 ,
\end{equation}
where $\clat\otto\clas$ is a point-wise product.
We also require that the homomorphism
\begin{equation}
\errep\ni t\mapsto \clat\in\cpdon
\end{equation}
be continuous (w.r.t.\ the the topology of
uniform convergence on compact sets on $\cpdon$).
Such semigroups can be classified. In fact, the symplectic
Fourier transform of a multiplication semigroup of functions of positive type on $\phasp$ is
a (continuous) convolution semigroup of probability measures, and convolution semigroups
on Lie groups admit a well known characterization associated with
the L\'evy-Kintchine formula~\cite{Heyer,Aniello1}.

Given a multiplication semigroup of functions of positive type
$\{\clat\}_{t\in\errep}\subset\cpdon$,
since $\clat$ is a bounded continuous function, we can define
a bounded operator $\hcsg_t$ in $\lrrn$:
\begin{equation} \label{setcsem}
\big(\hcsg_t \cinque f\big)(q,p):= \clat(q,p)\cinque f(q,p), \ \ \ f\in\lrrn.
\end{equation}
The set $\{\hcsg_t\}_{t\in\errep}$ is a semigroup of operators~\cite{Yosida-book}, namely,
\begin{enumerate}

\item $\hcsg_t\cinque \hcsg_s = \hcsg_{t+s}$, $t,s\ge 0$ \ (one-parameter semigroup property);

\item $\hcsg_0 = I$ \ (where $I$ is the identity operator).

\end{enumerate}

\begin{remark}
{\rm
Note that, since $|\clat(q,p)|\le\clat(0)=1$, $\{\hcsg_t\}_{t\in\errep}$ is a \emph{contraction semigroup}, i.e.,
$\|\hcsg_t\|\le 1$.
}
\end{remark}

\begin{remark}
{\rm
One may define an analogous contraction semigroup replacing the Hilbert space space $\lrrn$
with the Banach space $\lurrn$. The Fourier transform intertwines this further semigroup of operators
with a \emph{probability semigroup} in $\cz(\phasp)$, describing (under a certain assumption)
a drift-diffusion process~\cite{Aniello1,Aniello2,Aniello3}.
We leave the details to the reader.
}
\end{remark}

It is now natural to consider the restriction of the semigroup of operators $\{\hcsg_t\}_{t\in\errep}$
to a linear subspace of $\lrrn$. Indeed, as argued in sect.~\ref{posdef},
by complex linear superpositions one can extend the convex
cone $\qpdn$ of functions of quantum positive type on $\phasp$ to a dense
subspace $\lispn$ of $\lrrn$.
A semigroup of operators $\{\csg_t\}_{t\in\errep}$ in $\lispn$ is then defined as follows.
Since, according to Theorem~\ref{mainth}, the point-wise product of a continuous function of classical positive type by
a continuous function quantum positive type is a function of the latter type,
we can (consistently) set
\begin{equation} \label{setcsem-bis}
\big(\csg_t \cinque \qpt\big)(q,p):= \clat(q,p)\cinque \qpt(q,p), \ \ \ \qpt\in\lispn,
\end{equation}
where, with a slight abuse w.r.t.\ our previous notation, $\qpt$ here denotes a linear superposition of four functions
of quantum positive type:
\begin{equation}
\qpt = \qua_1 - \qua_2 + \ima\tre(\qua_3 - \qua_4), \ \ \
\qua_1,\ldots , \qua_4\in\qpdn \, .
\end{equation}
It is clear, moreover, that we have:
\begin{equation} \label{durel}
\csg_t \cinque \qpdn\subset\qpdn, \ \ \
\csg_t \cinque \qpdon\subset\qpdon.
\end{equation}

We will call the semigroups of operators $\{\hcsg_t\}_{t\in\errep}$ and $\{\csg_t\}_{t\in\errep}$
a \emph{classical-quantum semigroup} and a \emph{proper classical-quantum semigroup}, respectively.
Up to this point, the introduction of these semigroups of operators may be regarded
as a mere mathematical \emph{divertissement}, based on the properties of functions of
positive type. It turns out, however, that they do have a precise physical
interpretation which can now be unveiled by means of suitable quantization-dequantization maps.


\section{Introducing quantization-dequantization maps into the game}
\label{quan-dequa}

The procedure of associating with a quantum-mechanical operator (a state or an observable)
a function on phase space --- a Wigner function or a quantum characteristic function ---
may be thought of as the application of a suitable \emph{dequantization map}, which can be considered
as the reverse arrow of a \emph{quantization map} transforming functions into operators. The definition
of a \emph{star product}~\cite{AnielloFT,AnielloSP,Aniello-compact} --- the twisted convolution, in the case where
quantum characteristic functions are involved --- allows one to achieve a self-consistent formulation
of quantum mechanics in terms of phase-space functions and, as discussed in sect.~\ref{posdef}, a
fundamental step of this construction is the notion of function of quantum positive type. In order
to illustrate the meaning of this notion in terms of Hilbert space operators it is worth reconsidering
the dequantization map in a slightly more formal setting,
which also allows us to highlight its group-theoretical content.

A basic ingredient in the definition of a dequantization map is
a \emph{square integrable} (in general, projective) representation $U$ of a locally compact group $G$
in a Hilbert space $\hh$; see~\cite{Ali,Aniello-sdp,Aniello-sipr}, and references therein.
Here, $\hh$ should be though of as the Hilbert space of a quantum-mechanical system and $G$
as a symmetry group. Denoting by $\hs$ the Hilbert space of Hilbert-Schmidt operators in $\hh$,
the representation $U$ allows one to define a dequantization map
\begin{equation}
\dequm\colon\hs\rightarrow\ldg,
\end{equation}
see~\cite{Ali,AnielloFT,AnielloSP}, which is a (linear) \emph{isometry}.
In the case where the group $G$ is unimodular (e.g., a compact group) and
we consider, in particular, a trace class operator $\hrho$ in $\hs$,
the function $\dequm\tre\hrho$ associated with $\hrho$ is of the form
\begin{equation} \label{nonform}
(\dequm\tre\hrho)(g) =d_U^{-1}\,\tr(U(g)^\ast\hrho) ,
\end{equation}
where $\ddu>0$ is a normalization constant depending on $U$ and
on the choice of the Haar measure on $G$. The range $\ran(\dequm)$
of the dequantization map is a closed subspace of $\ldg$. The
quantization map associated with $U$ is the adjoint of the dequantization map, i.e.,
it is the partial isometry $\qum$ defined by
\begin{equation}
\qum:=\dequm^\ast\colon\ldg\rightarrow\hs.
\end{equation}
Clearly, we have that $\kerne(\qum)=\ran(\dequm)^\perp$, and in general
this kernel is not trivial. Also for the quantization map
one can provide explicit formulae; see, e.g.,~\cite{AnielloSP}. At this point, the star product
\begin{equation}
(\cdot)\stp(\cdot)\colon\ldg\times\ldg\rightarrow\ldg
\end{equation}
associated with $U$ is defined by
\begin{equation} \label{defstp}
f_1\stp f_2:= \dequm ((\qum f_1) \tre (\qum f_2)),
\end{equation}
with $(\qum f_1) \tre (\qum f_2)$ denoting the ordinary product (composition) of the
operators $\qum f_1$ and $\qum f_2$. Therefore, for a pair of functions living
in the range of the dequantization map the star product is nothing but the dequantized
version of the product of operators. Explicit formulae for star products
can be found in~\cite{AnielloSP}.

Let us consider the case where $G$ is the group of translations on
phase space $\phasp$~\cite{AnielloFT,AnielloSP}. In this case, $\hh=\ldrn$, $\ldg=\lrrn\equiv\lrrnco$
(the Haar measure is normalized in such a way that in~{(\ref{nonform})} $\ddu=1$) and the representation $U$ has to be
be identified with the \emph{Weyl system}~\cite{Aniello-WS,AnielloSP}, i.e.,
\begin{equation} \label{Weyl-sys}
U\qp =\disp,
\end{equation}
where $\hq=(\hq_1,\ldots,\hq_n)$, $\hp=(\hp_1,\ldots,\hp_n)$, with
$\hq_j$, $\hp_j$ denoting the $j$-th coordinate position and momentum operators in $\ldrn$.
The Weyl system is a (square integrable) projective representation,
\begin{equation}
U(q + \tq, p + \tp)= \emme(q,p\hspace{0.6mm};\tq,\tp)\sette U(q,p)\cinque U(\tq, \tp),
\end{equation}
where the multiplier $\emme$ is given by
\begin{equation} \label{multip}
\emme(q,p\hspace{0.6mm};\tq,\tp):=\exp\hspace{-0.5mm}\Big(\frac{\ima}{2}(q\cdot\tp-p\cdot\tq)\Big).
\end{equation}
It can be shown that for every density operator $\hrho$ in $\ldrn$, denoting with $\varrho$
the associated Wigner distribution, the function $\dequm\tre\hrho=\tr(U(g)^\ast\hrho)$ coincides with the
quantum characteristic function $\tvarrho$ defined by~{(\ref{wigetcha})}, see~\cite{Ali,AnielloFT,AnielloSP}.
Therefore, since $\fs=\fsy^\ast$, we have:
\begin{equation} \label{rewig}
\varrho = (2\pi)^{-n} \cinque \fs\tre\dequm\tre\hrho ;
\end{equation}
namely, quantization-dequantization \emph{\`a la} Weyl-Wigner
is obtained by composing the maps $\qum$ and $\dequm$ with the symplectic Fourier-Plancherel operator.

One can prove, moreover, the following facts~\cite{AnielloFT,AnielloSP}:
\begin{enumerate}

\item In this case, $\ran(\dequm)=\lrrn$, hence $\qum$ and $\dequm$ are unitary operators.

\item The quantization map $\qum$ intertwines the involution $\sinvo\colon \obse\mapsto \obse\invo$ in $\lrrn$, see~{(\ref{forinvo})},
with the standard involution $\opa\mapsto\opa^\ast$ in $\hsr$:
\begin{equation} \label{invint}
\qum \tre\sinvo\tre \obse = (\qum \cinque \obse)\invo .
\end{equation}

\item  The star product defined by~{(\ref{defstp})} coincides with the twisted convolution
$(\cdot)\starp(\cdot)$, see~{(\ref{deftc})}.

\end{enumerate}

From the previous facts and the results outlined in subsect.~{\ref{qcase}} one draws the
following further conclusions:
\begin{enumerate}
\setcounter{enumi}{3}

\item The quantum positivity condition~{(\ref{quapo})}, or~{(\ref{quapoalt})}, expressed
in terms of operators, becomes the condition that $\opb\in\hsr$ is such that
\begin{equation}
\tr(\opb\cinque\opa^\ast\opa)\ge 0, \ \ \ \forall\quattro\opa\in\hsr .
\end{equation}
This condition is equivalent to the fact that $\opb\ge 0$.

\item Denoting by $\trcr$ the Banach space of trace class operators in $\ldrn$, and by
$\trcrp$ the \emph{positive} trace class operators, we have that
\begin{equation}
\lispn=\dequm\sei\trcr, \ \ \ \qpdn=\dequm\sei\trcrp .
\end{equation}
Also note that, for every $\hrho\in\trcrp$, $\|\dequm\tre\hrho\tre\norinf=(\dequm\tre\hrho)(0) = \tr(\hrho)$.

\item Therefore, the quantization map $\qum =\dequm^\ast$ transforms bijectively
a function of quantum positive type
into a positive Hilbert-Schmidt operator, a \emph{continuous} function of quantum positive type
into a positive trace class operator and a \emph{normalized continuous} function of quantum positive type
into a density operator. By relation~{(\ref{invint})}, a function of quantum positive type $\qpt$
is selfadjoint: $\qpt=\sinvo\tre\qpt$.

\item Denoting by $\|\cdot\nordu$ the norm of $\lrrn$, with the normalization of the Haar measure
fixed above, for every density operator $\hrho$ in $\ldrn$ we have that
\begin{equation}
\|\dequm\tre\hrho\tre\nordu = \sqrt{\tr(\hrho^2)}=\sqrt{(\dequm\tre\hrho)\starp(\dequm\tre\hrho)(0)}\le 1;
\end{equation}
i.e., as anticipated in subsect.~\ref{qcase} (see~{(\ref{anticinor})}),
the norm of a normalized continuous function of quantum positive type,
regarded as a functional in $\lrrn$, corresponds to the square root of the purity of the associated state.

\end{enumerate}

We are finally ready to unveil the true nature of a classical-quantum semigroup,
see~{(\ref{setcsem})} and~{(\ref{setcsem-bis})}.
To this aim, first observe that associated with the Weyl system $U$ we have
a (strongly continuous) isometric representation $\urep$ of $\phasp$ acting in the Banach space
$\trcr$, i.e.,
\begin{equation}
\urep\qp\colon \trcr\ni\hrho\mapsto U\qp\sei\hrho \sei U\qp^\ast\in\trcr,
\end{equation}
which is the standard symmetry action of $\phasp$ on trace class operators.
Now, given a convolution semigroup $\{\mut\}_{t\in\errep}$ of probability
measures on $\phasp$, a semigroup of operators $\{\unimu\}_{t\in\errep}$ in $\trcr$
is defined by setting
\begin{equation} \label{deunim}
\unimu\sei\hrho := \intrrn  \urep(g)\sei\hrho\; \de \mut\qp .
\end{equation}
The integral of a vector-valued function on the rhs of~{(\ref{deunim})}
can be understood as a Bochner integral~\cite{Aniello2}. It can be shown that the
semigroup of operators $\{\unimu\}_{t\in\errep}$, a so-called \emph{twirling semigroup},
is actually a quantum dynamical semigroup (a completely positive, trace-preserving semigroup
of operators in $\trcr$); see~\cite{Aniello1,Aniello2,Aniello3,Aniello-OvF}.
\begin{theorem}
Let $\{\clat\}_{t\in\errep}$ be the multiplication semigroup of functions of positive type
associated, via the symplectic Fourier-Stieltjes transform, with the
convolution semigroup $\{\mut\}_{t\in\errep}$, i.e.,
\begin{equation}
\clat(q,p)=\intrrn \eee^{\ima (q\cdot p^\prime - p\cdot q^\prime)}\; \de\mut(q^\prime,p^\prime),
\end{equation}
and let $\{\csg_t\}_{t\in\errep}$ be the proper classical-quantum semigroup associated with
$\{\clat\}_{t\in\errep}$. Then, the quantization map $\qum$ intertwines
the semigroup of operators $\{\csg_t\}_{t\in\errep}$ with the quantum dynamical
semigroup $\{\unimu\}_{t\in\errep}$; namely, for every $\qpt\in\lispn$,
\begin{equation} \label{interqum}
\qum\dieci\csg_t \cinque\qpt = \unimu\sei\qum\cinque\qpt  , \ \ \ t\ge 0.
\end{equation}
\end{theorem}

\noindent {\it Proof:} The crucial observation is that $\dequm=\qum^\ast$
intertwines the representation $\urep$ with the representation $\two$
of $\phasp$ in $\lispn$ defined by
\begin{equation}
\big(\two(q,p)\hspace{0.4mm} \qpt\big)(\tq,\tp) :=
\ee^{-\ima(q\cdot\tp - p\cdot\tq)}\hspace{0.8mm} \qpt(\tq,\tp)   ,
\end{equation}
see~\cite{AnielloSP}. Relation~{(\ref{interqum})} then follows easily.~$\square$

We conclude observing that the isometric representation $\urep$
in $\trcr$ can be extended in a natural way to a unitary representation
in $\hsr$. This extension gives rise to a new semigroup of operators
in the Hilbert space $\hsr$, whose definition is completely analogous to~{(\ref{deunim})}.
This semigroup of operators is unitarily equivalent, via the
dequantization map, to the classical-quantum semigroup $\{\hcsg_t\}_{t\in\errep}$ in $\lrrn$.
We leave the details to the reader.


\section{Conclusions and perspectives}
\label{conclus}

Continuous functions of positive type on phase space play a remarkable role in classical statistical mechanics and
in the WWGM formulation of quantum mechanics. In the classical setting, they can be defined as
positive functionals on the group algebra $\big(\lurrn, \convo, \cinvo\big)$, and continuity can be regarded
as a byproduct of this definition. In the quantum setting, continuous functions of (quantum) positive type
are embedded in the positive functionals on the twisted convolution $\Hstar$-algebra
$\big(\lrrn, \starp, \sinvo\big)$. This corresponds, via quantization, to the embedding of positive
trace class operators in the positive Hilbert-Schmidt operators.
We have therefore found it natural to consider these functionals,
implemented by square integrable functions, as a quantum analogue of functions of (classical) positive type.
The functions representing quantum states are characterized in this framework precisely by the property of being continuous
and by a suitable normalization condition, which in general differs from their normalization as functionals.

Note that, following~\cite{Folland-AA} --- where (standard) functions of positive type are discussed in detail ---
we have made a precise distinction between \emph{functions of positive type} and \emph{positive definite functions},
both in the classical and in the quantum setting. The two notions coincide if the functions involved are continuous.

Analyzing the fundamental properties of functions of classical and quantum positive type it is also
natural to consider a class of semigroup of operators, the \emph{classical-quantum semigroups}.
We have recently introduced these semigroups of operators in~\cite{AnielloCQS}, with the aim
of highlighting their relations with other classes of semigroups of operators and
the role that they play in quantum information science. As shown in sect.~\ref{quan-dequa},
by quantizing a classical-quantum semigroup one obtains a quantum dynamical semigroup
belonging to the class of \emph{twirling semigroups}. Every twirling semigroup is generated by a pair
formed by a projective representation $U$ of a locally compact group and by a convolution semigroup $\{\mut\}_{t\in\errep}$
of probability measures on that group~\cite{Aniello1,Aniello2,Aniello3}. In the case where $U$ is a Weyl system,
and $\{\mut\}_{t\in\errep}$ is a \emph{Gaussian} convolution semigroup~\cite{Aniello1,Aniello3,Heyer},
the associated twirling semigroup consists of certain quantum dynamical maps called \emph{classical-noise channels}
in the context of quantum information, see~\cite{Giovannetti} and references therein.
Therefore, borrowing this terminology  we can conclude that, under a suitable assumption, by quantizing
a classical-quantum semigroup one obtains a \emph{classical-noise semigroup}.

Here we have focused on the group-theoretical background of classical-quantum semigroups;
in particular, on their relation with functions of positive type.
This study should pave the way to extend the results obtained
for the group of translations on phase space to other groups. We believe that
a basic ingredient for achieving this extension should be a suitable generalization of the notion
of function of positive type on a locally compact group, and that in this regard
previous results on group-theoretical star products~\cite{AnielloSP} should
be a useful starting point.


\section*{Acknowledgments}

The author wishes to dedicate this paper, with admiration and gratitude,
to M.A.~Man'ko and V.I.~Man'ko, shining examples of scientists.
The author also wishes to thank the organizers of the international conference
\emph{Operator and Geometric Analysis on Quantum Theory} (14-19 September 2014, Levico Terme, Italy),
where the main results of the paper were presented, for their kind hospitality.


\end{document}